\documentclass[showpacs,amssymb,twocolumn,aps]{revtex4}
\usepackage{amsmath}
\usepackage{graphicx} 
\usepackage{epsf}
\usepackage{graphics}
\usepackage[latin1]{inputenc}
\newcommand{\be}{\begin{equation}}
\newcommand{\ee}{\end{equation}}
\newcommand{\bea}{\begin{eqnarray}}
\newcommand{\eea}{\end{eqnarray}}

\newcommand{\vp}{|\vec p|}

\begin{document}

\title{Thermal Effective Lagrangian of Static Gravitational Fields}

\author{F. T. Brandt  
and J. B. Siqueira} 
\affiliation{ Instituto de F\'{\i}sica,
Universidade de S\~ao Paulo,
S\~ao Paulo, SP 05315-970, Brazil}

\begin{abstract}
We compute the effective Lagrangian of static gravitational fields
interacting with thermal fields. Our approach employs the usual
imaginary time formalism as well as the equivalence between the static and space-time independent
external gravitational fields. This allows to obtain a closed form
expression for the thermal effective Lagrangian in $d$ space-time dimensions.
\end{abstract}


\maketitle

\section{Introduction}

The high temperature limit in thermal field theory has many
interesting properties which, in some cases, allow to obtain closed
form expressions for quantities like the effective Lagrangians in gauge
theories \cite{Braaten:1992gm}. When the gravitational interactions are
taken into account, there are also indications that it may be possible
to obtain an effective Lagrangian, though until now this has not been
demonstrated in general. Only in the { static limit}, when the
fields are time independent, it is possible to shown that all the
one-loop thermal Green functions can be generated from
an effective Lagrangian which has a simple closed form \cite{Rebhan:1991yr}.
%

More recently, it has been shown that the static limit of thermal
Greens functions in gauge theories coincide with the limit when all the external four
momenta are equal to zero. 
This has been explicitly verified for individual thermal Green functions
 \cite{Frenkel:2009pi} (in another work the long wavelength limit has
 also been investigated \cite{Brandt:2009ht}). This result indicates that
in configuration space we may make the hypothesis that a static
background is equivalent to a space-time independent configuration, 
in the high temperature limit.
The purpose of the present paper is to further investigate this
issue in the context of a gravitational background,
using a more general approach which can be easily extended to all orders.
In this paper, we say that the background metric is  ``static''  
when it does not depend on time, which is less restrictive than the condition of
a ``static space-time'' when, additionally,  $g_{0i}=0$, in some refference frame.

At one-loop order, the effective Lagrangian of the gravitational
fields interacting with thermal scalar fields can be written in terms
of the functional determinant as follows 
\be\label{eq1}
{\cal L} = \frac{T}{V} \log \left[ {\rm Det}\left(-\beta^2 p_\mu \sqrt{-g}g^{\mu\nu}  p_\nu \right) \right]^{-1/2}_{T},
\ee
where $p_\mu = -i \partial_\mu$, $g=(-1)^d \det{g_{\mu\nu}}$ and $d$ is the space-time dimension.
This expression is based on the usual approach
which is employed in order to obtain the one-loop effective Lagrangians in field theory \cite{Dunne:2007rt}.
Here we are considering that the temperature $T=1/\beta$ is much bigger than
any other mass scale, such as the scalar field mass, and the subscript
$T$ is to  make explicit that we are considering only the temperature
dependent part of the determinant. We will employ the
imaginary time formalism \cite{kapusta:book89,lebellac:book96,das:book97}. 

In the next section we will consider a perturbative method which
express ${\cal L}$ in terms of powers of the gravitational field
$\tilde h^{\mu\nu}$ in a Minkowski background. The purpose of the perturbative calculation is to make contact with
some known results in the static limit.
In section III we derive the effective Lagrangian for a general static background. As a check, we verify that
the perturbative results of section II can be obtained from the exact result of section III. 
Finally, in section IV we discuss the results and perspectives.
Some details of the perturbative calculation are left to the appendix.

\section{Fields in a Minkowski background}

In order to make contact with some known perturbative results,
we define the gravitational field $\tilde h^{\mu\nu}$ as \cite{Goldberg:1958aa}
\be\label{eq2}
\sqrt{-g} g^{\mu\nu} \equiv \tilde g^{\mu\nu} = \eta^{\mu\nu} + \tilde h^{\mu\nu},
\ee
where $\eta^{\mu\nu}$ is the Minkowski metric. Here we have a symmetric tensor field $\tilde h^{\mu\nu}$, in
an Minkowski background.
Inserting Eq. \eqref{eq2} into Eq. \eqref{eq1}, yields
\be\label{eq3}
{\cal L} = \frac{T}{V} 
\log \left[ {\rm Det}( -\beta^2 p^2)  {\rm Det}\left(1+\frac{1}{p^2} p_\mu  
\tilde h^{\mu\nu} p_\nu \right) \right]^{-1/2}_{T}.
\ee
We now make use of the hypothesis that the zero momentum limit can give us information about the
static limit. This makes it possible to write
\begin{eqnarray}\label{eq4}
{\cal L}^{\rm stat.} &=& T \log \left[ {\rm Det}( -\beta^2 p^2)  {\rm Det}\left(1+\tilde h^{\mu\nu} \frac{p_\mu p_\nu}{p^2} \right) \right]^{-1/2}_{T}.
\nonumber \\
&=& {\cal L}^{(0)}-\frac{1}{2}\frac{T}{V} 
\log {\rm Det}\left(1+\tilde h^{\mu\nu} \frac{p_\mu p_\nu}{p^2}\right), 
\end{eqnarray}
where ${\cal L}^{(0)}$ is given by
\cite{kapusta:book89,lebellac:book96,das:book97}
\be\label{eq5}
{\cal L}^{(0)} = \frac{\Gamma[d] \zeta(d)}{2^{d-2}
  \pi^{(d-1)/2}\Gamma\left(\frac{d-1}{2}\right)  (d-1)} T^d .
\ee 
$\Gamma$ is the Gamma function and $\zeta$ is the Riemann zeta
function. Eq. \eqref{eq5} is simply the Stefan-Boltzmann pressure of the free
gas in $d$ space-time dimensions, so that the second term in Eq. \eqref{eq4} can also be
interpreted as the corrections to the pressure due to the interation
with the gravitational field.

Let us now consider the quantity
\be\label{eq9}
{\cal L}^I = -\frac{1}{2} \frac{T}{V} 
\log {\rm Det}\left(1+\tilde h^{\mu\nu} \frac{p_\mu p_\nu}{p^2}\right).
\ee
Using the relation $\log {\rm Det} A = {\rm Tr}\log A$, the imaginary time formalism leads to
\be\label{eq10}
{\cal L}^I = -\frac{T}{2} \sum_{n=-\infty}^{\infty} \int
\frac{d^{d-1} p}{(2\pi)^{d-1}}
\log\left(1+\tilde h^{\mu\nu} \frac{p_\mu p_\nu}{p^2}\right),
\ee
where $p_0= i \omega_n = i 2\pi n T$ and $p^2 = \eta^{\mu\nu} p_\mu p_\nu 
= p_0^2 - \vp^2 = -(2\pi n T)^2 - \vp^2$. It is understood that we are using the
reference frame where the heat bath is at rest.

Let us now investigate the properties of Eq. \eqref{eq9} employing
a perturbative expansion in powers of $\tilde h^{\mu\nu}$. Upon using the expansion
$\log(1+x) = x - x^2/2 + x^3/3 + \cdots $, Eq. \eqref{eq9} can be
written as
\begin{eqnarray}\label{eq11}
{\cal L}^I &=& -\frac{T}{2} \sum_{n=-\infty}^{\infty} \int \frac{d^{d-1} p}{(2\pi)^{d-1}}
\nonumber \\
&\times&\left(\tilde h^{\mu\nu} \frac{p_\mu p_\nu}{p^2}
-\frac{\tilde h^{\mu\nu} \tilde h^{\alpha\beta}}{2} 
\frac{p_\mu p_\nu p_\alpha p_\beta}{p^4} + \cdots \right)
\nonumber \\&=&
\tilde h^{\mu\nu} I_{\mu\nu} + \frac{1}{2}\tilde h^{\mu\nu} \tilde h^{\alpha\beta} I_{\mu\nu\alpha\beta} + \cdots ,
\end{eqnarray}
where 
\be\label{eq12}
I_{\mu\nu} = -\frac{T}{2}
\sum_{n=-\infty}^{\infty} \int \frac{d^{d-1} p}{(2\pi)^{d-1}}
\frac{p_\mu p_\nu}{p^2}
\ee
and
\be\label{eq13}
I_{\mu\nu\alpha\beta} = \frac{T}{2}
\sum_{n=-\infty}^{\infty} \int \frac{d^{d-1} p}{(2\pi)^{d-1}}
\frac{p_\mu p_\nu p_\alpha p_\beta}{p^4}.
\ee
Each individual term in Eq. \eqref{eq11}  is promptly identified as the $n$-point one-loop Feynman diagram with vanishing
external momentum, contracted with $n$ fields $\tilde h^{\mu\nu}$. The calculation of the first two terms in Eq. \eqref{eq11} can be done
in a straightforward way. The details of this calculation is presented in the appendix.

Combining the  Eqs. \eqref{eq4}, \eqref{eq11}, \eqref{eq14}, \eqref{eq18}, \eqref{eq21} and \eqref{eq22ab}, 
the effective Lagrangian can be written as follows
\be\label{eq27}
{\cal L}^{\rm stat.} =  {\cal L}^{(0)} + \tilde \Gamma_{\mu\nu} \tilde h^{\mu\nu} 
+ \frac{1}{2} \tilde \Pi_{\mu\nu\alpha\beta} \tilde h^{\mu\nu} \tilde h^{\alpha\beta} 
+ \cdots ,
\ee
where
\be\label{eq28}
\tilde \Gamma_{\mu\nu} = \frac{{\cal L}^{(0)}}{2} \left(d u_\mu u_\nu - \eta_{\mu\nu}\right)
\ee
and
\begin{eqnarray}\label{eq29}
\tilde \Pi_{\mu\nu\alpha\beta} & = & {\cal L}^{(0)} \left[
\tilde \Gamma_{\mu\nu} \tilde \Gamma_{\alpha\beta} + \tilde \Gamma_{\mu\alpha}\tilde \Gamma_{\nu\beta} + \tilde \Gamma_{\mu\beta} \tilde \Gamma_{\nu\alpha}
\right. \nonumber \\
&-& \left.\frac{d(d-1)}{2} u_\mu u_\nu u_\alpha u_\beta \right] .
\end{eqnarray}
Both results in Eqs. \eqref{eq28} and  \eqref{eq29} are exactly the same as one would obtain for the 
static limit of the one-loop Feynman diagrams.

One can verify that there are Weyl identities which relates 
$\tilde \Pi_{\mu\nu\alpha\beta}$ with $\tilde \Gamma_{\mu\nu}$. Indeed,
\begin{subequations}\label{eq30}
\begin{eqnarray}
\eta^{\mu\nu} \tilde \Gamma_{\mu\nu} & = & 0 ,
\\
\eta^{\mu\nu} \tilde \Pi_{\mu\nu\alpha\beta} & = & -\tilde \Gamma_{\mu\nu} .
\end{eqnarray}
\end{subequations}
These identities are a consequence of the conformal symmetry under the transformation 
$\sqrt{-g} g^{\mu\nu} \rightarrow (1+\epsilon) \sqrt{-g} g^{\mu\nu}$, 
which is equivalent to $\tilde h^{\mu\nu} \rightarrow
\tilde h^{\mu\nu} + \epsilon \tilde h^{\mu\nu} +
\epsilon \eta^{\mu\nu}$, with $\epsilon$ infinitesimal. 
Even in the non-static case it is known that the Weyl identities are satisfied
by the high temperature thermal amplitudes \cite{Brandt:1993bj}.
This is an important information which constrains the general form of the Lagrangian
and may help to obtain a closed form in terms of the exact metric tensor, 
although this has not been achieved yet.

One can also show that ${\cal L}^{\rm stat}$ is independent of the representation of the
graviton field. As an example, alternatively one could define the graviton field as
\be\label{eq31}
\bar h^{\mu\nu} = g^{\mu\nu} - \eta^{\mu\nu}.
\ee
The relation between $\tilde h^{\mu\nu}$ and $\bar h^{\mu\nu}$ can be
readily found to be 
\begin{eqnarray}\label{eq32}
\tilde h^{\mu\nu} &=& \bar h^{\mu\nu} - 
\eta^{\mu\nu}\left(\frac{\bar h}{2}-\frac{1}{4} \bar h_{\alpha\beta} \bar h^{\beta\alpha} -\frac{\bar h^2}{8}\right) 
\nonumber \\ &-&  \frac{\bar h}{2} \bar h^{\mu\nu}  + {\cal O}(\bar h^3),
\end{eqnarray}
where the raising and lowering of indices are performed with the Minkowski metric.
Inserting Eq. \eqref{eq32} into Eq. \eqref{eq27} and using the Weyl
identities in Eq. \eqref{eq30} one can see that ${\cal L}^{\rm stat.}$ has the same form
when the graviton field is defined as $\bar h^{\mu\nu}$. This independence on the graviton field 
parametrization is expected for a physical quantity like the effective Lagrangian even
in more general cases, when the field transformation is not induced by a simple rescaling of the metric,
as in the previous example. In general, both the fields and the amplitudes
would change in such a way to preserve the invariance of the Lagrangian \cite{Brandt:1993bj}.
For instance, transforming to the graviton representation $h_{\mu\nu} = g_{\mu\nu} - \eta_{\mu\nu}$, we obtain
\be\label{eq27a}
{\cal L}^{\rm stat.} =  {\cal L}^{(0)} +  \Gamma^{\mu\nu}  h_{\mu\nu} 
+ \frac{1}{2}  \Pi^{\mu\nu\alpha\beta}  h_{\mu\nu}  h_{\alpha\beta} 
+ \cdots ,
\ee
where
\be\label{eq28a}
\Gamma^{\mu\nu} = -\frac{{\cal L}^{(0)}}{2}\left(d u^\mu u^\nu - \eta^{\mu\nu}\right)
\ee
and
\begin{eqnarray}\label{eq29a}
\Pi^{\mu\nu\alpha\beta} & = & {\cal L}^{(0)} \left[
\Gamma^{\mu\nu} \Gamma^{\alpha\beta} -\frac{1}{4}\left(
\eta^{\mu\alpha}\eta^{\nu\beta} + \eta^{\mu\beta}\eta^{\nu\alpha}
\right)
\right. \nonumber \\
&+& \left.\frac{d}{2} u^\mu u^\nu u^\alpha u^\beta \right]
\end{eqnarray}
which are both in agreement with the static result obtained in \cite{Rebhan:1991yr}.
Also, $\Gamma^{\mu\nu}$ and $\Pi^{\mu\nu\alpha\beta}$ satisfy the identities \eqref{eq30}.
We remark that while the result in Eq. \eqref{eq29a} exhibits only the Bosonic symmetry and the symmetry 
associated with the  metric, Eq. \eqref{eq29} has a larger tensor symmetry, as could be anticipated from
Eq. \eqref{eq13}. 

\section{General static backgrounds}
Let us now consider a more general physical scenario described by a background
metric which is not necessarily close to the Minkowski metric. Using again the relation 
$\log {\rm Det} A = {\rm Tr}\log A$ in the context of the imaginary time formalism, Eq. \eqref{eq1} yields
\be\label{eqe1}
{\cal L} = -\frac{T}{2} \sum_{n} \int \frac{d^{d-1} p}{(2\pi)^{d-1}} 
\log\left(-\beta^2 p_\mu \tilde g^{\mu\nu}  p_\nu \right).
\ee
Using the hypothesis of a space-time independent metric, the Lagrangian can be written as follows
\begin{eqnarray}\label{new1}
{\cal L}^{\rm stat.} &=& -\frac{T}{2} \sum_{n} \int \frac{d^{d-1} p}{(2\pi)^{d-1}} 
\nonumber \\ 
& \times & \log\left[-\beta^2 \left(
g^{00} p_0^2 + 2 g^{0i} p_0 p_i + g^{ij} p_i p_j \right) \right],
\end{eqnarray}
where we have split the metric in its space-time components and neglected terms which are temperature independent.

Let us now perform the change of variables
\be\label{trans1}
p_i \rightarrow p_i^\prime = M_i^{j} p_j + f_i p_0,
\ee
where $M$ is symmetric. Upon imposing the condition
\be
p_i^\prime p_i^\prime = g^{ij} p_i p_j + 2 g^{0i} p_0 p_i + f_i f_i p_0^2,
\ee
we obtain
\be\label{new2}
\left\{
\begin{array}{lll}
M_i^{j} M_i^{k} &=& g^{jk}
\\
f^{i} M^{ij} &=& g^{0j}
\end{array}
\right. .
\ee
Therefore, the effective Lagrangian can be written as
\begin{eqnarray}\label{new3}
{\cal L}^{\rm stat.} &=& -\frac{T}{2} \frac{1}{\sqrt{-{\det}{{\bf g}}}}\sum_{n} \int \frac{d^{d-1} p^\prime}{(2\pi)^{d-1}} 
\nonumber \\ 
& \times & \log\left[-\beta^2 \left(
(g^{00} - f^i f^i)p_0^2 - p_i^\prime p_j^\prime \right) \right],
\end{eqnarray}
where the entries of the matrix ${\bf g}$ are $g^{ij}$. Performing the transformation 
\be\label{trans2}
p_i\rightarrow \sqrt{g^{00} - f^j f^j} \;p_i
\ee
 we readily obtain
\be\label{new4}
{\cal L}^{\rm stat.} = {\cal L}^{(0)}
\frac{\left(g^{00} - ({\bf g}^{-1})^{ij} g^{0i} g^{0j} \right)^{\frac{d-1}{2}}}{\sqrt{-{\det \bf g}}}.
\ee
A straightforward calculation shows that the perturbative results obtained in the 
previous section can be generated from the Lagrangian  \eqref{new4}. Indeed, using Eq. \eqref{eq2}, the
first order contribution from Eq. \eqref{new4} is simply
\be
{\cal L}^{(1)} = \frac{{\cal L}^{(0)}}{2}  \left(
d \tilde h^{00} -\tilde h \right),
\ee
which is the same as the first order term in \eqref{eq27}. 
Proceeding similarly, the second order contribution from \eqref{new4} produces
\begin{eqnarray}\label{new5}
{\cal L}^{(2)} &=& \frac{1}{8} {\cal L}^{(0)}  \left[d(d-2) (\tilde h^{00})^2 - 2 d \tilde h^{00} \tilde h  
 +\tilde h^2 
\right. \nonumber \\
&+&  \left.  2\left(\tilde h_{\mu\nu} \tilde h^{\mu\nu}\right)_{\tilde h_{0i}=0} - 4 (d-1) \tilde h_{0i} \tilde h^{0i}\right],
\end{eqnarray}
which is also in agreement with the second order contribution from Eq. \eqref{eq27}.
We recall that, as a consequence of conformal invariance, the results are  
unchanged under the transformation $g^{\mu\nu} \rightarrow \tilde g^{\mu\nu} \equiv \sqrt{-g} g^{\mu\nu}$.

We can also express the exact result in terms of the co-variant metric components. Using the
identity  (this follows from $g^{\mu\alpha} g_{\alpha \nu} = \delta^\mu_\nu$) 
\be\label{iden1}
g^{00} - ({\bf g}^{-1})^{ij} g^{0i} g^{0j} = (g_{00})^{-1},
\ee
Eq, \eqref{new4} can be written as
\be\label{eqe4a}
{\cal L}^{\rm stat.} = {\cal L}^{(0)} 
\frac{\sqrt{-g_{00} \det {\bf g}^{-1}}}{(g_{00})^{d/2}} ,
\ee
Expanding the determinant of $g^{\mu\nu}$ in terms of co-factors and using the identity
$
g_{i0} = - {\bf g}^{-1}_{ij} g^{j0} g_{00}
$
as well as \eqref{iden1} we can show that  
\be
g_{00} \det {\bf g}^{-1} = \frac{1}{\det{g^{\mu\nu}}} = g. 
\ee
Therefore, Eq. \eqref{eqe4a} yields
\be\label{eqe4}
{\cal L}^{\rm stat.} = {\cal L}^{(0)} 
\frac{\sqrt{-g}}{(g_{00})^{d/2}} ,
\ee
which is in agreement with the known result when $d=4$ \cite{Rebhan:1991yr}.
This static effective Lagrangian can also be obtained using a much more involved approach in terms of the
heat-kernel technique restricted to a static space-time, in a refference frame such that $g_{0i} = 0$ \cite{Alwis:1995cr}. 
Since the heat bath breaks the invariance under general coordinate transformations, as it is
evident  due to  the presence of the Matsubara sum in Eq. \eqref{eqe1}, it is essential to perform
the calculation for general values of $g_{0i}$. Physicaly one must impose that the heat bath is freely 
moving in a time-like geodesic, so that in the heat bath frame $g_{0i}$ vanishes only in very special cases, even for static
space-times.

\section{Discussion}

In this paper we have presented a simple method which allows to obtain
the effective Lagrangian of static gravitational fields interacting with thermal fields. 
A key ingredient in this analysis was the hypothesis that the effective Lagrangian 
can be obtained from a space-time independent background. Conversely,
one can also claim, using the perturbative results of section II, 
further support to this hypothesis.

It is not very difficult to extend the present analysis to include spinor and gauge thermal fields. 
A straightforward calculation shows that the only modification is the replacement of ${\cal L}^{(0)}$ 
by the corresponding free contributions of fermions or gauge bosons.

We point out that the perturbative expansion of the static 
effective Lagrangian given by Eq. \eqref{eqe4}, is in agreement with the known 
static limit of Feynman amplitudes. On the other hand, we have demonstrated in this paper that 
\eqref{eqe4} can be directly derived from the functional determinant given by Eq. \eqref{eq1}, when
the metric is space-time independent. This is consistent with the
analytical behavior
of individual thermal Feynman amplitudes \cite{Frenkel:2009pi}. Therefore, 
the present approach may be a suitable starting point towards the analysis of more general backgrounds. 
This would be useful in order to obtain  a closed form expression for other 
background configurations, such as 
the long wavelength limit, or even more general gravitational backgrounds, using the known symmetries which
are characteristic of the high temperature limit.

The result given by Eq. \eqref{eqe4} may also be viewed as the pressure of a weakly interacting gas 
subjected to an external gravitational field. An obvious extension of the present analysis would be 
to consider the contributions which are higher than one-loop, so that the effects of interactions 
between the gas particles would be taken into account, and more realistic applications could be considered. 
This may be interesting in the context of stellar evolution or cosmology.


\appendix

\section{}
In this appendix we compute the first and second order terms in Eq. \eqref{eq11}.
Let us first notice that the space-time trace of $I_{\mu\nu}$ in Eq. \eqref{eq12}, 
\be\label{eq13a}
\eta_{\mu\nu} I^{\mu\nu} = -\frac{T}{2}
\sum_{n=-\infty}^{\infty} \int \frac{d^{d-1} p}{(2\pi)^{d-1}},
\ee
can be set equal to zero in the context of the dimensional dimensional regularization technique.
Therefore, the most general form of $I_{\mu\nu}$ can be written as
\be\label{eq14}
I_{\mu\nu} = a\left( d u_\mu u_\nu - \eta_{\mu\nu} \right),
\ee
where $u_\mu$ is such that $\eta^{\mu\nu} u_\mu u_\nu = 1$ and can be
identified with the heat bath four-velocity (in the rest frame of the
heat bath $u_\mu = (1, 0, 0, \cdots, 0)$). 
Projecting both sides of Eq. \eqref{eq14} along the tensor 
$\left( d u_\mu u_\nu - \eta_{\mu\nu} \right)$ and using
Eq. \eqref{eq12} yields
\be\label{eq15}
a = -\frac{T}{2}\frac{1}{d-1}
\sum_{n=-\infty}^{\infty} \int \frac{d^{d-1} p}{(2\pi)^{d-1}}
\frac{p_0^2}{p^2}.
\ee 
Using $p_0^2 = p^2 + \vp^2$ and the dimensional regularization
prescription for the temperature independent terms, 
\be\label{eq16}
a = \frac{1}{2}\frac{1}{d-1}
\int \frac{d \Omega}{(2\pi)^{d}} \int_0^\infty \vp^d S^1(\vp) d \vp,
\ee 
where
\begin{eqnarray}\label{eq17}
S^1(\vp) & = & T\sum_{n=-\infty}^{\infty}  \frac{1}{(2\pi n T)^2+\vp^2}
\nonumber \\
& = & \frac{1}{2\vp} + \frac{1}{\vp}\frac{1}{{\rm e}^{\vp/T} -1}.
\end{eqnarray}
Substituting Eq. \eqref{eq17} into Eq. \eqref{eq16} and using again the
dimensional regularization prescription for the $T$-independent term,
we finally obtain
\be\label{eq18}
a = \frac{1}{2}\frac{1}{d-1}
\int \frac{d \Omega}{(2\pi)^{d}} \int_0^\infty  
\frac{\vp^{d-1}}{{\rm e}^{\vp/T} -1} d\vp = \frac{{\cal L}^{(0)}}{2},
\ee
where in the previous expression we have identified the result for the effective Lagrangian of free scalar fields.
From Eqs. \eqref{eq4}, 
\eqref{eq11},  
\eqref{eq14} and \eqref{eq18} we obtain
the following result up to the first order
\begin{eqnarray}\label{eq19}
{\cal L}^{\rm stat.} & = & {\cal L}^{(0)}\left[1 + \frac{1}{2} \left(d u_\mu u_\nu -
    \eta_{\mu\nu}\right)\tilde h^{\mu\nu} \right]
+ \cdots
\nonumber \\ & = & 
{\cal L}^{(0)}\left[1 + \frac{1}{2} \left(d \tilde h^{00} - \tilde h \right)\right]
+ \cdots .
\end{eqnarray}
In the second line of the previous equation the effective Lagrangian is explicitly expressed in the rest frame of the thermal bath.
 
Let us now consider  the second order contribution to ${\cal L}^{\rm stat}$.
From the tensorial symmetry of the Eq. \eqref{eq13} one can see that the result 
of the integration can be written in terms of three independent tensors as follows
\begin{eqnarray}\label{eq21}
I^{\mu\nu\alpha\beta} & = & b_1 u^\mu u^\nu u^\alpha u^\beta
\nonumber \\ & + & 
b_2\left( \eta^{\mu \nu} u^\alpha u^\beta + \mbox{symm.}\right)
\nonumber \\ & + & 
b_3\left( \eta^{\mu \nu} \eta^{\alpha \beta} + \mbox{symm.}\right).
\end{eqnarray}
Projecting both sides of Eq. \eqref{eq21} along each of the three
tensors and solving the resulting system, yields
\begin{subequations}\label{eq22}
\begin{eqnarray}
b_1 & = & -(d+2) {\cal L}^{(0)} + (d+2)(d+4) b \\
b_2 & = & \frac{{\cal L}^{(0)}}{2} - (d+2)b \\
b_3 & = & b ,
\end{eqnarray}
\end{subequations}
where
\be\label{eq23}
b = \frac{1}{2}\frac{1}{(d-1)(d+1)}
\int \frac{d \Omega}{(2\pi)^{d}} \int_0^\infty \vp^{d+2} S^2(\vp) d \vp ,
\ee
and
\be\label{eq24}
S^2(\vp) =  T\sum_{n=-\infty}^{\infty} \frac{1}{\left[(2\pi n T)^2+\vp^2\right]^2}  
\ee
Using the identity
\be\label{eq25}
\frac{1}{\left[(2\pi n T)^2+\vp^2\right]^2}  =
-\frac{1}{2\vp}\frac{d}{d\vp}\frac{1}{(2\pi n T)^2+\vp^2}
\ee
and performing integration by parts, one can proceed similarly to the
first order calculation, yielding the result
\be\label{eq26}
b = \frac{{\cal L}^{(0)}}{4}.
\ee
Substituting Eq. \eqref{eq26} into Eq. \eqref{eq22}  we obtain the following result for the three
structure constants in Eq. \eqref{eq21}
\begin{subequations}\label{eq22ab}
\begin{eqnarray}
b_1 & = & \frac{d(d+2)}{4} {\cal L}^{(0)}  \\
b_2 & = & -\frac{d}{4} {\cal L}^{(0)} \\
b_3 & = & \frac{1}{4} {\cal L}^{(0)}.
\end{eqnarray}
\end{subequations}

\acknowledgments

F. T. Brandt and J. B. Siqueira would like to thank J. Frenkel and D. G. C. McKeon for helpful discussions and
CNPq for financial support.


\end{document}